# Further considerations of cosmic ray modulation of infra-red radiation in the atmosphere

K.L. Aplin(a*) and M. Lockwood(b)
(a) Department of Physics, University of Oxford, Keble Road, Oxford OX1 3RH UK
(b) Department of Meteorology, University of Reading, PO Box 243, Earley Gate, Reading RG6 6BB UK
*corresponding author

**Abstract**

Understanding effects of ionisation in the lower atmosphere is a new interdisciplinary area, crossing the traditionally distinct scientific boundaries between astro-particle and atmospheric physics and also requiring understanding of both heliospheric and magnetospheric influences on cosmic rays. Following the paper of Erlykin et al.,[1] we develop further the interpretation of our observed changes in long-wave (LW) radiation[2], by taking account of both cosmic ray ionisation yields and atmospheric radiative transfer.  To demonstrate this, we show that the thermal structure of the whole atmosphere needs to be considered along with the vertical profile of ionisation.  Allowing for, in particular, ionisation by all components of a cosmic ray shower and not just by the muons, reveals that the effect we have detected is certainly not inconsistent with laboratory observations of the LW absorption cross section.  The analysis presented here, although very different from that of Erlykin et al., does come to the same conclusion that the events detected by AL were not caused by individual cosmic ray primaries – not because it is impossible on energetic grounds, but because events of the required energy are too infrequent for the 12 hr$^{-1}$ rate at which they were seen by the AL experiment.  The present paper numerically models the effect of three different scenario changes to the primary GCR spectrum which all reproduce the required magnitude of the effect observed by AL. However, they cannot solely explain the observed delay in the peak effect which, if confirmed, would appear to open up a whole new and interesting area in the study of water oligomers and their effects on LW radiation.   We argue that a technical artefact in the AL experiment is highly unlikely and that our initial observations merit both a more wide-ranging follow-up experiment and more rigorous, self-consistent, three-dimensional radiative transfer modelling.

1. *Introduction*

In 2013 Aplin and Lockwood[2] (hereafter AL) reported observations of a small change in atmospheric infra-red (IR, also here referred to a "longwave", LW) absorption in a narrow band around 9 μm, associated with high energy particle events detected by a galactic cosmic ray (GCR) telescope. AL attributed their finding to IR absorption from molecular cluster-ions (MCI, sometimes referred to as "small ions") that form rapidly after ionisation due to acquisition of hydrogen-bonded molecules such as water.  The IR absorption results from the energy absorbed by the bending and stretching of the hydrogen bonds within the cluster. IR absorption is a well-known property of polar atmospheric molecules, although the IR absorption properties of both water vapour and clusters remain poorly understood.  Atmospheric molecular cluster-ions are created by secondary atmospheric particles generated by a primary GCR ionising the column of air above their experiment. In a follow-up paper in this journal, Erlykin et al.[1] (hereafter ESW) raised questions related to AL's discovery.

Interdisciplinary work frequently provokes challenges in interpretation, and with this in mind, the issues raised by ESW are addressed here.

ESW refer to the AL observations as a "large absorption" and call them a "remarkable result", which we believe gives an entirely false impression. The longwave radiation detected by AL ("downward longwave" or "DLW") is mainly radiated from the Earth and then returned to the surface having been absorbed by greenhouse gases and re-radiated back down from the atmosphere[3]. The resulting broadband DLW flux for the AL experiment averaged 324 Wm$^{-2}$, consistent with averages of other observations[3,4]. No detectable GCR signal was found in this broadband measurement, but there was evidence of a response in the 9μm wavelength band studied, after coincident detection of a GCR secondary particle by the cosmic ray telescope. This narrow band was selected because laboratory experiments had revealed that, within it, MCI could absorb the LW radiation[5,6] and AL were investigating if any such effect could be detected in the atmosphere from the ground using the DLW radiation. The peak of the median effect seen by AL across many events was found to be about 6mWm$^{-2}$ for the lower quartile of the simultaneously detected broadband downward heat fluxes measured. Since these fluxes were less than 295 Wm$^{-2}$, with a minimum DLW of 231 Wm$^{-2}$, the signal is, at a maximum, 0.003% of the broadband DLW. Hence, in terms of the modulation of atmospheric radiative balance, it is self-evident that the effect can only be very modest. From a rough calculation, AL inferred a centennial-scale change in radiative forcing from this effect of between −1 and −10 mWm$^{-2}$. (Note that it is negative, meaning its effect is a long-term cooling contribution). This is very small compared to other known factors: for example the change in trace greenhouse gas concentrations between pre-industrial times and 2005 gives about 2.6 Wm$^{-2}$ and the estimated change in total solar shortwave (SW) irradiance gives a radiative forcing of about 0.2 Wm$^{-2}$.[7,8] The radiative forcings due to other individual greenhouse gases are much larger (of order 1.66 Wm$^{-2}$ for $CO_2$, 0.48 Wm$^{-2}$ for $NH_4$ and 0.16 Wm$^{-2}$ for $N_2O$)[9]. It is therefore unclear how the effect, as reported by AL, can be described as "large".

ESW deduce that the cosmic ray primaries creating muons that enter the AL telescope have the moderate energy of 12GeV. This is not inconsistent with the AL telescope count rates. However, ESW then state that "the average multiplicity of secondary particles at this primary energy will be of order 10", and they try to show that the triggering events produce so few secondary ionising particles (and then only along the tracks of the few muons) that the MCI absorption cross-section must be unphysically high. A better explanation is that, although the AL triggering particles are almost certainly muons, muons are not the main source of the atmospheric ionisation causing much of the small IR fluctuations seen after the trigger event. Indeed, in the context of longwave radiative transfer, muon-induced MCI are likely to be of lesser significance. Usoskin and Kovaltsov[10] have demonstrated how for low energy primaries (200 MeV), the ionisation is almost entirely due to hadrons. Ionisation induced by high-energy cosmic rays (100 GeV) is dominated by muons in the lower troposphere and by the electromagnetic component (electrons, positrons, photons) in the mid-troposphere upwards. For middle energies (10GeV) electromagnetic ionisation dominates in the upper troposphere/lower stratosphere (UTLS) region, hadronic in most of the rest of the troposphere, and muons only very close to the surface. One reason why ESW find only very small absorption is because they consider the effect of 10 secondary muons and do not allow for the electromagnetic and hadron cascades – apart from there being relatively few muons, the ionisation they produce only dominates in the boundary layer, whereas it is well known that the ionisation maximum occurs at the Pfotzer-Regener maximum of ionisation at around 15 km altitude[11,12]. The integrated columnar ionisation is usually dominated by the electromagnetic cascade. In this paper we consider ionisation from all three sources, hadron, electromagnetic and muon.

In discussing the implications of the AL experiment for radiative changes caused by air ions it is important to realise that there are two completely different situations that must be considered and which ESW confuse. The AL experiment detected an effect following individual transient ionising events (of some kind) that triggered the cosmic-ray detector. They detected an average transient response to these relatively rare events (12 per hour) in which they infer that air ions are formed and subsequently decay away: these events end in a return to steady-state conditions. AL never suggested that all cosmic rays would have the same effect as the trigger events: indeed they specifically knew that this was not the case because otherwise the continuous GCR precipitation would have given a constant layer of ionisation (as is observed in the atmosphere[12]) and so give a constant effect on the DLW rather than the transient response that was observed. In this paper we study the likely characteristics of the cosmic rays that trigger responses in both the cosmic ray detector and the narrowband DLW detector. In contrast to these transient events, the radiative forcing calculations must relate to the steady-state situation where the production (from all sources) and loss of the air ions (and of their potentially IR-active products) are in balance. ESW arrive at an absurd radiative result by assuming that every cosmic ray incident on the atmosphere contributes the same IR absorption effect seen, on average, following one of AL's triggering events. They also assume that the integrated effect is a simple accumulation, such that increasing the rate of events by a factor N would cause N times the effect on the absorption of downwelling longwave radiation. This is invalid because the steady-state situation relevant to the radiative profiles is not the sum of the rare transient effects and arises from a balance between the ion production and loss rates.

In their final sentence, ESW state that the AL data should be analysed with respect to randomised triggers: this is of course appropriate and was, in fact, precisely the approach that was taken by AL to generate their results, as described in their section 3.1. The conclusion drawn by ESW, that the results seen are most likely caused by cross-talk between the telescope and radiometers was also already addressed in some detail by AL (see their section 2). It is reiterated here that a lag of 20s between the triggering event and the first radiometer measurement was used, both to avoid any instantaneous cosmic ray effects in electronics or the detector contributing to the IR measurement, and to give the radiometer adequate time to respond to any IR changes. In addition, tests in both the laboratory and at the field site never once displayed any symptoms of the cross-talk ESW attribute the AL results to. It is therefore highly unlikely that the AL findings are caused either by cross-talk or random fluctuations.

However, all this is not to say that ESW do not raise some valid issues, in particular in relation to the nature of the events that trigger the events, as will be discussed in this paper. This does have implications for AL's estimate of a $10^{12}$ energy amplification factor and for their interpretation of the delay in peak DLW response in terms of long MCI lifetime and spatially-localised ionisation enhancements that drift into the relevant part of the radiometer field of view.

2. *General Considerations*

Figure 1 shows some altitude profiles needed to explain the effect discussed by AL. Ionisation is generated in the atmosphere by the precipitation of GCRs. The continuous flux of GCRs of all energies, E, reaching the top of the atmosphere and a wide range of zenith angles $\chi$ yields a horizontally-stratified layer of MCI. The solid line in figure 1(a) shows modelled ionisation rate profile, q(h), for a time when the heliospheric shielding effect on the GCRs reaching the top of Earth's atmosphere is quantified by a "modulation potential"[10] of $\phi$ = 270 kV (which applies for relatively low solar activity and is an average for the interval of the AL experiment[13]). The modelling will be outlined in more detail in section 4.1. The dashed line shows a profile observed during a balloon flight on 23 May 2013,[11] when the heliospheric modulation potential was $\phi$ = 679 kV (moderate solar activity[13]) which has been scaled to allow for the dependence on $\phi$, as predicted by the model. The solid line in figure 1(b) shows the ion concentration $N_i$ derived for steady-state balance between this production rate q(h) and the estimated loss rate, and the dashed line shows the results from a balloon flight on 15 May 1979,[12] (when $\phi$ = 706 kV, again normalised to $\phi$ = 270 kV using the model dependence[13]). As in other tests of the model,[10,11,14,15] figure 1 shows close agreement with the observations. We note two significant areas of disagreement between the modelled and observed $N_i$ profiles. The first is at low altitudes, where the observed $N_i$ is lower than predicted: this may be in part because the observed q in figure 1(a) is also lower at these altitudes, but probably also points to a role of aerosols in giving enhanced loss of ionisation (see section 4.1). The other area of disagreement is at the highest altitudes where observed $N_i$ becomes considerably larger than predicted. This is almost certainly due to uncertainties in the ion-ion recombination coefficient used but is not of concern here because (as demonstrated by figure 2b) the DLW flux at such altitudes is negligible.

Figure 1(c) shows a typical DLW flux profile observed during a balloon flight[4]. This profile has a characteristic form because greenhouse gases absorb outgoing terrestrial LW radiation and then re-radiate it (both upward, returning a fraction to the OLW, and downward to give the DLW) according to their temperature at that height. Because this re-radiation follows Planck's law we can estimate the part of it that lies within the narrowband filter used in the experiment of AL, centred on the wavelength of $\lambda$ = 9.15μm with a width (FWHM) of 0.9μm. The fraction of the Planck spectrum emitted by the surface that is returned to Earth within the band depends on the atmospheric greenhouse gases active at those wavelengths, largely ozone (as the experiment band is in the tail of the nearby broad ozone line at 9.3-10.1μm) and water vapour which give an atmospheric transmission at 9.15μm of about 80%[16]. At the surface, the total broadband DLW is near 330 Wm$^{-2}$ and we find 6% ($\approx$20 Wm$^{-2}$) of this lies in the total passband of AL's narrowband radiometer.

Note that in addition to this terrestrial DLW there is a small amount of IR power in the long-wavelength tail of the spectrum of solar ("shortwave", SW) radiation incident on the Earth (the solar SW and terrestrial surface LW powers are equal at approximately $\lambda$ of 4μm and the terrestrial DLW dominates at longer wavelengths). In the narrowband radiometer band at $\lambda$ = 9.15μm, solar SW gives about 0.2 Wm$^{-2}$ which is just 1% of the atmospherically re-radiated terrestrial DLW and is neglected here.

The profiles shown in figure 1 are key to understanding the effect of MCI on DLW. Because the peak of the $N_i$ profile is at altitudes at which the DLW is very small, this part of the $N_i$ profile has a relatively small effect on the DLW seen at the ground. On the other hand, the rapid increase in DLW with decreasing altitude makes MCI at altitudes below the $N_i$ peak of greater importance: such ions will be preferentially generated by the more energetic part of the primary GCR spectrum.

In this paper, we do not attempt full radiative transfer analysis[17] which would self-consistently allow for the effect of any additional absorption (as postulated by AL to be caused by additional MCI generated in one of the events) on the temperature profile of the atmosphere. Rather we adopt a typical DLW profile and assume it is not perturbed by the additional LW absorption.

2. *Effect of a single primary GCR*

The interpretation presented by AL was in terms of the additional ionisation generated in events caused by a single energetic primary GCR which generated ionisation on top of the pre-existing steady-state profile exemplified in figure 1(b). Following ESW's comments we re-analyse this interpretation.

The right hand side of figure 2 is a schematic that gives several of the geometrical parameters needed to describe the AL experiment. The detectors (the cosmic ray telescope, the broadband radiometer and the narrowband radiometer) are all situated at O. An individual GCR primary is shown arriving at a zenith angle of $\chi$ such that its path, extrapolated to the Earth's surface, would reach the point B which is a distance d from O. The point A is at a height $h_o$ and is where the concentration of MCI generated by the GCR begins to increase rapidly as the GCR descends, as shown by the schematic ionisation profile due to the primary GCR shown on the left. A is taken to be the apex of the ionisation cone which has a half angle $\beta_C$. Using a cone of ionisation is a major difference to ESW's approach (which was to consider MCI to only be formed along the tracks of the few muons generated): one of the roles of the electromagnetic component is to spread the ionisation into a cone[18]. The centre of the hadron, muon and electromagnetic cones, and hence of the total $N_i$ cone, would be the line AB. The ionisation cone causes MCI to appear in the $2\pi$ steradian upward field of view of the radiometers. To rigorously and self-consistently compute the effect on the downwelling long wave radiation in this case would entail a full three-dimensional radiative transfer analysis, as for example has been carried out to study the LW absorption effect of aircraft contrails[17]. This would require consideration of the three dimensional distribution of MCI in the field of view and integration over all possible values of the elevation and azimuth angles ($\varepsilon$ and $\propto$, respectively). However, the contrails example demonstrates that we can gain a first-order insight into the instrument response with a one-dimensional analysis of the LW flux vertically down ($\varepsilon = \pi/2$) and the effect of the ionisation formed vertically above O.

We have calculated the DLW absorption for various energy GCR primaries and cone angle and find the maximum effect is always for particles precipitating down the vertical above O, i.e. for d = 0 and $\chi$ = 0. This is not surprising as it places the maximum number of MCI generated in the path of the LW to the detector. The energy required to generate an ion pair is $\Delta E$ = 35eV and hence the maximum ionisation yield of a E = 35 GeV primary is $E/\Delta E = 10^9$ ion pairs (the limit in which all the primary energy is dissipated in the production of ionisation) i.e. it generates a total of $\Sigma = 2 \times 10^9$ ions. Ions will have more effect on DLW at lower altitudes because the DLW flux profile shown in figure 1c. In order to estimate a maximum effect on DLW we take the top of the ionisation cone to be at an altitude $h_o$ = 10 km and the diameter of the ionisation cone on the ground to be 10m for which the cone angle $\beta_C$ = 0.06° and the volume of the cone is $V_C = (\pi/3) h_o^3 \tan^2(\beta_C) \approx 10^6 m^3$. Hence the mean ion concentration in the cone for a 35GeV primary is $<N_i> = \Sigma/V_C \approx 2\times10^3 m^{-3}$. For the optimum geometry with d = 0 and $\chi$ = 0, this gives $\int N_i dh = <N_i>h_o \approx 2\times10^7$ ion $m^{-2}$. The total absorption of DLW seen at the ground is given by

$$\Delta F_{DLW} = \int_o^{h_o} F_{DLW}(h) \sigma N_i(h) dh \qquad (1)$$

$F_{DLW}(h)$ is the DLW flux at height h and $\sigma \approx 10^{-15} m^2$ is the absorption cross section. The part of the DLW spectrum that is in the experiment band is about 20 $Wm^{-2}$ at the surface and has a profile similar to that of the broadband power shown in figure 1(c). If we simplify by taking $F_{DLW}(h)$ to be constant at its average value over the altitude range 0-10 km of $<F_{DLW}> \approx 10 Wm^{-2}$ in the waveband of the experiment, we obtain $\Delta F_{DLW} \approx <F_{DLW}>\sigma <N_i>h_o \approx 2\times10^{-7} Wm^{-2}$. This maximum estimate for the additional effect of a single 35 GeV primary is smaller than the average of the peak effect detected in the AL experiment of $4\times10^{-3} Wm^{-2}$ by a factor $2\times10^4$. Hence the effect seen by AL could only be explained by a single GCR primary of energy larger than the 35 GeV employed in this calculation by the same factor (so the total ion yield is $\Sigma = 4 \times 10^{13}$), i.e. with an energy exceeding 7 PeV. This is close to the "knee" of the cosmic ray spectrum and we would expect to see such events at the rate of one every few months, rather than the 12 per hour detected in the AL experiment.

Hence although this calculation is very different from ESW's, using a realistic downward longwave radiation flux profile and allowing for ionisations by all components (muon, hadron and electromagnetic), we nevertheless do agree with their conclusion that the small absorption cross section $\sigma$ means that the events observed by AL are not generated by single GCR primaries, the concept that was used in the original interpretation by AL. The problem is not that a single GCR is incapable of producing the required ionisation, rather that the events observed by AL are far too frequent in their occurrence for this to be a possibility.

3. *The LW absorption cross section, $\sigma$*

In their paper, ESW use a rough estimate of $\sigma = 2\times10^{-15}$ m$^2$ for the absorption cross section of the 9$\mu$m LW absorption wavelength used in the AL experiment. This value is taken from the laboratory experiments[6] and is of the correct order of magnitude but is actually slightly high when one considers the bandwidth of the AL experiment. We here use a Gaussian approximation to the AL narrowband filter response function $f_d(\lambda)$ with characteristics of a central wavelength of 9.15$\mu$m and width (FWHM) of 0.9$\mu$m. The response is actually achieved using two filters and the full $f_d(\lambda)$ for the instrument is given in the bottom panel of figure 2 of [19]. In addition to the main line around 9.15$\mu$m, the instrument does some have sidebands with discrete lines at 14$\mu$m and 15.2$\mu$m and a broad response between about 19$\mu$m and 22.5$\mu$m which is a plateau at almost exactly half the response of the main line at 9.15$\mu$m. These sidebands may have some significance. In terms of spectral wavenumber the main line is at 1042-1149 cm$^{-1}$, the two discrete sidebands are at 714 cm$^{-1}$ and 658 cm$^{-1}$ and the plateau is at 444-526 cm$^{-1}$.

The top panel of figure 3 shows the MCI spectral line seen in the laboratory measurements of the fractional absorption, A as a function of wavelength, $\lambda$.[6] The middle panel shows the detector response $f_d(\lambda)$ and the bottom panel the observed absorption weighted by the filter response function, A$\times f_d(\lambda)$. It can be seen the band used by AL is somewhat wider than the absorption line. Averaging over the filter bandwidth the weighted mean response to the line is $< Af_d(\lambda)> = 0.69\%$. Hence $\Delta F/F = 0.0069 = \sigma\int N_i dl$, where dl is an element of the path length, and $N_i$ the ion concentration. The best estimate of $\int N_i dl$ for the laboratory experiment was $10^{13}$ m$^{-2}$, which gives a $\sigma$ value of $0.69\times10^{-15}$ m$^2$. The measurement uncertainties were estimated to be 50% and hence our best estimate of $\sigma$ for the experiment narrow band is $(0.7\pm0.35)\times10^{-15}$ m$^2$.

4. *Effects of GCR spectrum changes*

ESW dismiss the AL results as "cross-talk" between the instruments, but in extensive tests of the combined radiometers and cosmic ray telescope experiment, both in the laboratory and at the field site, this effect has never been seen to occur, nor does it explain the nature of the response. We therefore remain convinced that there is a genuine physical explanation of AL's results. Section 3 shows that AL's interpretation of the event trigger being a single primary GCR is inconsistent with the event occurrence frequency. In this section we pursue three potential alternative explanations.

    4.1 *Model calculations*

To compute the absorbed DLW flux we employ equation (1) with modelled profiles of $N_i$ based on the atmospheric ionisation yield model of Usoskin and Kovaltsov.[10] We follow their numerical recipe (outlined below) with one minor modification to the way allowance is made for the shielding effect of the geomagnetic field on low-energy GCRs. The series of equations that define this model starts from fixed forms for the flux of the local interstellar spectrum (LIS) of GCRs of a given species, $J_{LIS}(E)$ (expressed in units of (GeV/nucleon)$^{-1}$m$^{-2}$sr$^{-1}$s$^{-1}$) and then allows for the shielding effect of the heliosphere using so-called "force field model" equations which quantify the shielding effect using the modulation potential $\phi$ to derive the spectrum outside Earth's magnetosphere, $J_i(E)$ for each primary species. The model then computes the vertical geomagnetic cut-off rigidity $P_C$ for the GCRs at the relevant geomagnetic latitude (which is the same for all GCR species), that corresponds to the cut-off energy $E_C$ (which depends on the GCR rest mass). In the original model formulation, a sharp cut-off at $E_C$ was applied, such that the spectral density $J_i(E)$ is multiplied by a factor $f_{co}(E)$ which is zero for $E < E_C$ and unity for $E \geq E_C$. However, this neglects the variation in the geomagnetic cut-off energy with zenith angle and that the quoted cut-off applies to the vertical direction but is lower for large zenith angles to the west and higher to the east.[20] To allow for this we use the less sharp spectral cut off provided by an empirical fit to observations for the function $f_{co}(E)$:

$$f_{co}(E) = (1 + (E_C/E)^{12})^{-1} \qquad (3)$$

For every energy E the modelled ionisation yield for a given GCR species $Y_i(E,x)$ is then computed at each atmospheric depth, x (the mass of air in the column above the height considered), by interpolation of the tables in [10]. We convert x to altitude h using the density profile measured during the balloon flight that yielded the DLW profile adopted[4]. From this the ionisation rate q is computed by integrating the ionising effect of the whole spectrum

$$q(h) = \Sigma_i \int_o^\infty Y_i(E,h) \, J_i(E) \, f_{co}(E) \, dE \qquad (4)$$

where the sum is over all primary species (the dominant two species, protons and alphas, are considered here). Equation (4) differs from that of Usoskin and Kovaltsov[10] in that the lower limit of integration is zero rather than $E_C$ because it includes the term $f_{co}(E)$ to allow for the geomagnetic cut-off.

To compute the mean ion concentration $N_i$ we need to consider the continuity equation and hence ion loss. The ions are lost by direct recombination or by ion-aerosol attachment,[21,22] so that

$$dN_i/dt = q - \alpha N_i^2 - \beta Z N_i \qquad (5)$$

where $\alpha$ is the ion-ion recombination coefficient (generally taken to be $1.6 \times 10^{-12}$ m$^3$ s$^{-1}$ at Earth's surface[22]) $\beta$ is the ion-aerosol attachment coefficient (a complex function of aerosol radius and charge),[21] and Z is the aerosol concentration. The steady-state solution ($dN_i/dt = 0$) to equation (5) is:

$$N_i = \{ (r^2 + 4q/\alpha)^{1/2} - r \}/2 \qquad (6)$$

where $r = \beta Z/\alpha$. It must be remembered that q, $\alpha$, $\beta$, Z, r, and hence $N_i$, are all functions of altitude h. Because $\beta$ is a complex function of ion mass, aerosol particle radius and charge and the aerosol concentration profile is variable and unknown, we here take the approach of taking clean air (Z = 0) and making a first order allowance for aerosols by changing $\alpha$ to an effective recombination coefficient[22] and, in particular, we adopt the effective $\alpha$ profile presented in figure 4 of Rosen and Hofmann[23]. This is the procedure used to compute the ionisation rate and steady-state ion concentration profiles shown in parts (a) and (b), respectively, of figure 1 for a site at geomagnetic latitude of 52° (rigidity cut off $P_C$ = 2GV) and a heliospheric modulation potential of $\phi$ = 270 kV.

To deal with time-dependent situations, we commence from these steady state profiles but then vary the GCR primary in a prescribed way starting from this steady state. We then evolve the profile numerically using equation (5) at each height. At every time step we compute the DLW absorption seen at the surface $\Delta F_{DLW}$ (in the narrowband of the AL experiment) using equation (1).
For the initial steady state profile, $\Delta F_{DLW}$ = 336.9 mWm$^{-2}$. Note that the estimated uncertainty in experiment and model comparisons of DLW[24] is of order $\pm 2$ Wm$^{-2}$ and hence the value derived here is only about 17% of this uncertainty. Hence the DLW effect predicted for the laboratory cross sections is a very minor component and well within the uncertainties of our understanding of DLW. Thus the effect is certainly not remarkably large, as ESW concluded.

The profile of DLW absorption per unit height, $dF_{DLW}/dh$, is given in figure 4(a). The structure seen at the lowest altitudes is an artefact of the numerical interpolation used to predict the ionisation yield functions at large optical depths. This structure is also seen in figures 1(a) and 1(b) but is amplified when, as in figure 4, an altitude gradient is taken. The profile represents the combined effect of the altitude profiles in both the DLW flux and the ion concentration. Figure 4(a) shows that the absorption is roughly constant at about $3.5 \times 10^{-7}$ Wm$^{-3}$ up to an altitude of about h = 7 km, but then decays away almost exponentially.

### 4.2 *Primary GCR spectrum changes*

We modulate the GCR primary spectrum at the top of the atmosphere according to an event timeseries that is discussed in the next sub-section. The peak change seen during this timeseries is determined for three different scenarios described here:

A. There is an increase in the spectral density of GCRs $J_i(E)$ reaching the magnetosphere at middle energies. We here consider a 6% increase in the energy range 10 < E < 100 GeV. On a log-log plot of $J_i(E)$ as a function of E this is an imperceptible change. Such a change could arise from localised structure in the inner heliosphere or even from a fluctuation in the local interstellar flux. This is not unreasonable because observations of primary GCRs in different energy ranges, for example from the BESS-Polar I experiment, indicate that this energy band can show temporal changes of a similar magnitude to the changes we assume, that are not seen at lower energies[24].

B. There is a 20% decrease in the effective heliospheric modulation parameter, $\phi$ caused by heliospheric structure. Greater magnitude increases are seen on the timescale of hours ahead of Forbush decreases, caused by the passage of transient solar wind structures such as coronal mass ejections, co-rotating interaction regions and current sheet crossings[25,26]. We are not aware of any observational evidence showing what magnitude of changes are possible on minute timescales.

C. There is a 20% decrease in the geomagnetic cut-off rigidity caused by changes in the geomagnetic field. This corresponds to the effect of a 2° change in the geomagnetic latitude of the station. Changes corresponding to 5° are regularly detected during geomagnetic storms[27] but again we are not aware of any information on what magnitude of changes can be caused by more rapid fluctuations in the geomagnetic field.

The amplitudes of the effects have been chosen because an iterative study reveals they all give roughly the same peak effect on the DLW at the surface. Of the three scenarios, we regard A as the most likely. Figure 4(b) shows how each change perturbs the altitude of the rate of DLW absorption at its peak effect. Figure 5 analyses the effect that these three changes have on the ionisation rate profile, q(h). For scenario A, q at h = 50 km is increased by 0.4% which rises with decreasing height to 3.4% at the surface. For scenario B the fractional increase in q falls with height over the same range from 7% to 1% and for scenario C it falls from 28% to 0.3%. Scenario A has a such a large relative effect on DLW as it generates additional ionisation at low altitudes (where DLW is largest) because of the increase in energetic primary fluxes. On the other hand, a relatively large change in geomagnetic rigidity cut-off is needed (scenario C) as it allows a greater flux of low energy particles to reach the top of the atmosphere and these have a preferential effect at higher altitudes where DLW is low. In scenario B, the change in ϕ increases the flux at all energies, but affects lower energies more than higher ones. Hence the fractional change again increases with height, but less so than for scenario C.

Ionisation rates from balloon flights show short-term (minute scale) fluctuations in q(h) of order 20%, rising to about 100% at the lowest altitudes[11]. Much of this variability is due to the limited counting statistics of the detectors (particularly at the lowest altitudes) but real changes in flux of the scale shown in figure 5 are certainly possible, particularly for scenarios A and B.

Figure 4(b) shows the effect of these changes on the DLW absorption profile shown in figure 4(a). In each case, the increase in ($dF_{DLW}/dh$) relative to that for the steady state case, $\Delta(dF_{DLW}/dh)$, is shown using the same line types as in figure 5. The plot shows that the largest additional absorption for scenario A is introduced at the surface but there is considerable contribution from greater heights with an almost linear decrease up to about h = 15 km. For scenario B, the effect peaks around h = 7km and for scenario C it peaks around h = 9 km, with a larger contribution at greater altitudes than for the other cases.

### 4.3 *Temporal waveform of the change*

To evaluate the time-dependent changes in the ionisation profile we need a realistic waveform for the imposed changes. (The scenarios given in the previous section are for the changes at their peak). To obtain this we look at the distribution of the intervals between coincident triggers of the two Geiger tubes in AL's cosmic ray telescope. This distribution is inherent in the plot given in the bottom panel of their figure 3 because it is a second trigger which brings to an end each superposed data series in the composite. Allowing for the latency of the device in the seconds following a trigger and using polynomial fitting to smooth the data, we obtain the probability of a coincident GCR detection, $P_{GCR}$, as a function of time over AL's experiment shown by the upper panel of figure 6. The low count rates of the detectors[30] mean that this waveform is not seen in each event, but figure 6(a) gives an average for all the events in the AL experiment dataset. We now study the implications of this average waveform. At the peak of the event (simulation time 600s), when an event trigger is most likely, the full percentage change described in the previous section is applied (hence only at this time do the ionisation rate change profiles shown in figure 4 apply) at other times the change in the GCR spectrum is equal to $P_{GCR}$ times the peak effect. Note that at simulation time t = 0, $P_{GCR}$ = 0 and the ionisation rate and concentration profiles shown in figures 1(a) and 1(b) apply.

### 4.4 *Simulated variations of the narrowband DLW power*

Panel (b) of figure 6 shows the derived DLW absorbed power as a function of time in these simulations for the three scenarios. It can be seen the effect is very similar in all three cases (remember that the peak amplitude in each scenario has been iterated to make this the case). The peak effect is seen roughly 80s after the peak of the GCR spectrum change. The subsequent decay is similar in all three cases. Because the three scenarios generate additional ionisation preferentially at different heights, the recombination rate would vary (because the recombination coefficient $\alpha$ is a function of altitude) and from the heights of the peak effects on the DLW shown in figure 4(b) we might expect this to cause differences in the response and recovery relaxation times. However, the effect seen on the ground is the integral of the profiles shown in figure 4(b) and the effects have been scaled to give roughly the same amplitude of effect on the ground. This, along with the relative constancy of $\alpha$ at low h, causes the net recovery time constant to be rather similar in all three cases.

The thick solid line shows the average effect reported by AL. It can be seen that all three scenarios reproduce the amplitude of the effect detected by AL. However they cannot reproduce the time delay of the observed response. This is because the response time for the ion concentration, $N_i$ is $\tau$

= 1/($\alpha N_i$) and if $N_i$ is large enough to cause detectable DLW absorption, $\tau$ for the expected $\alpha$ becomes small. As pointed out by AL, at the surface $\alpha \approx 1.6 \times 10^{-12}$ m$^3$s$^{-1}$ and the modelled $N_i \approx$ 1$\times 10^9$ m$^{-3}$ giving $\tau \approx$ 625s which is consistent with AL's observed lag. However, at h = 7km, $\alpha \approx 2 \times 10^{-12}$ m$^3$s$^{-1}$ and $N_i \approx 4 \times 10^9$ m$^{-3}$ giving $\tau \approx$ 125 s which is more consistent with the average lag shown in figure 6(b) (of order 80s). On the other hand, the observational composite requires a lag of about 500s on average. Hence the delay in peak response detected by AL does not appear to be caused by the timeconstant for the decay in the MCI produced, as was invoked by AL. This point is discussed further in the following section.

5. *Discussion*

In the above sections we have shown that although the additional DLW absorption detected by AL could be caused by individual energetic primary GCRs, the rate of such events would be much lower than the 12 hr$^{-1}$ of the AL experiment. We have studied the effect of three scenarios for rapid (several minute timescale) changes in the GCR spectrum and found what amplitude of such changes is required to have an effect on the DLW absorption of the observed magnitude. Of the three scenarios, we think A is the most likely and is the most consistent with the variability of ionisation rate seen in balloon flights on these short timescales.

However, the model shows that, for the additional ion concentrations required, the MCI recombination timescales do not appear to be consistent with the observed delay in the AL experiment. From the above, the predicted timeconstants are too short by a factor of order 500/80 $\approx$ 6. We now discuss possible causes of this.

The most obvious solution to this problem is to increase the energies at which the $J_i(E)$ spectrum is enhanced. If more of the additional ionisation is produced closer to the surface, its the ions needed to cause the additional absorption would be at lower altitudes and of lower $N_i$ which would increase the decay timeconstant. However, we believe that there are other possibilities that may be more important.

It is possible that the effective recombination coefficient $\alpha$ we have used is too large. This could be because if the MCI are very massive their thermal speed at a given temperature is reduced which reduces the probability of ion-ion recombination occurring. The theory by J.J. Thomson predicts that for charged particles the recombination coefficient $\alpha$ varies as the root mean square of the positive and negative ions' thermal speeds[28] and hence will vary as the inverse square root of their mass. Hence a factor 6 could be achieved if their mass was 36 times greater than expected. This seems a large factor, and hence a less probable cause.

Potentially more interesting is the possibility that electrically-enhanced recombination does not stop the recombined clusters from acting as IR absorbers. Indeed the laboratory work of Carlon specifically finds evidence that this is the case[5]. There is an equilibrium between atmospheric charged and neutral clusters, which can be formed by dissociation or recombination, and all of which are expected to be active IR absorbers [28,29]. The recombined water clusters are electrically-neutral water oligomers consisting of, or containing, $(H_2O)_n$ for which n would decline as they evaporate – the sequence ending with tetramers (n=4), trimers (n =3) and dimers (n=2) before returning to the monomer, the water molecule (n= 1). There has been a great deal of interest in the IR absorption spectra of these water oligomers, particularly at wave numbers between about 3000 and 3500 cm$^{-1}$ ($\lambda$ = 2.9 - 3.3$\mu$m) where the stretching vibration of the O-H bond gives a considerable IR absorption. Information on the effects of large clusters is sparse, but experiments by Goss et al.[30] are of interest as they relate to a mix of clusters with n between 10 and 100 and covering a wider-than-usual range of wavenumbers (4000-700 cm$^{-1}$, $\lambda$ = 2.5–14.8$\mu$m). These reveal that, as well as the shorter O-H wavelength bond stretching effect, at the long wavelength end of this spectrum there is IR absorption caused by intermolecular vibrations. The Goss et al. experiment did not show any increased amplitude in the main 9μm absorption band of the AL experiment, but these and other results support absorption of water clusters in our experiment's sidebands[30,31] This offers a potential explanation of the longer-than-expected decay time constant, and even of the continued rise in absorption as large n clusters split into multiple clusters of smaller n. There is also some evidence that some of the better-studied smaller oligomers, such as the tetramer, do have absorption lines in the main band of the AL experiment[32]. The time constants for these neutral cluster changes are predicted to be of order several hundreds of seconds[28] and hence this does provide a real possibility of explaining the delay before peak DLW absorption.

The analysis presented in this paper raises many new questions beyond those of ESW. These new considerations include the limitation that nature of the inferred change in the GCR primary spectrum is not known and the existence, decay and infrared absorption of the postulated oligomers in the experiment passband are largely undetected in the atmosphere. The reason that so little is known about this area is that the necessary experiments have not been made; but they never will be if interesting indications are dismissed as instrumental effects. Another problem now clearly evident is that this is indeed a highly interdisciplinary area with elements of interstellar cosmic ray physics, heliospheric physics, magnetospheric physics, atmospheric chemistry and radiation transfer physics. ESW state that further work is necessary, firstly, to repeat and understand the atmospheric effect observed by AL. Secondly, more collaborative work between particle and atmospheric physicists is required to overcome the interdisciplinary barriers that clearly still exist between the two communities. We agree completely with both these points, but not because we think the events seen by AL are simply spurious cross-talk between two instruments (a possibility that instrument tests have eliminated) but because the consequences of the MCI absorption of LW have not been explored at all in the ever more important context of the atmosphere and Earth's radiation budget. The AL experiment was a test exploration, and hence the experiment was run for only a short time – enough only to get a statistically significant sample of trigger events. During this interval there was little variation in the incident cosmic ray fluxes. Observations during a major Forbush decrease would have been very interesting and from the analysis presented here should have given a detectable imprint on the DLW.

One exciting possibility would be to place broadband and narrowband DLW radiometers, or atmospheric spectrometers close to state-of-the-art energetic cosmic ray instrumentation, along with lower-energy GCR instruments such as neutron monitors and ionisation chambers. Much of this infrastructure already exists at the Pierre Auger Observatory[33,34]. Our study also highlights the need to understand the full altitude profile of both the longwave fluxes and the ionisation and this requires instrumented balloon flights, perhaps using modified meteorological radiosondes carrying ionisation detectors[11], with the addition of aerosol and radiative measurements.

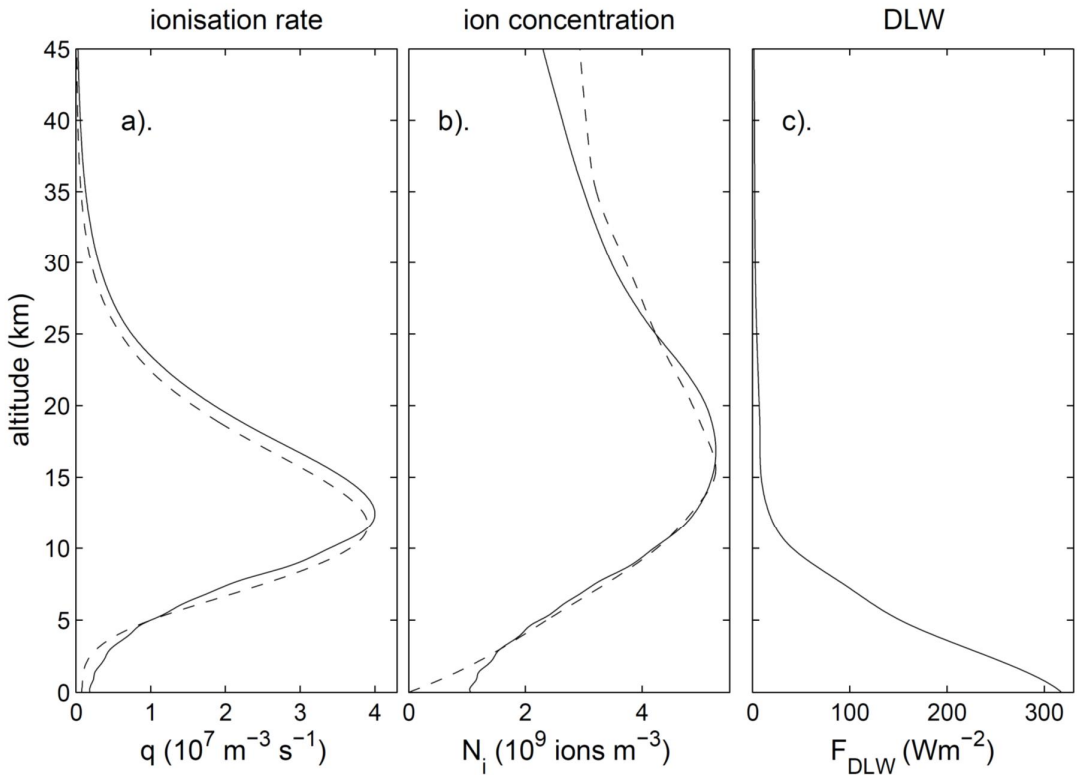

**Figure 1.** Atmospheric altitude profiles of: (a) cosmic ray ionisation rate, $q(h)$; (b) steady-state ion concentration, $N_i(h)$ and (c) downward longwave flux (DLW), $F_{DLW}(h)$. The solid lines in (a) and (b) were computed using the ionisation yield functions of Usoskin and Kovaltsov[10] for a heliospheric modulation potential $\phi = 270$ MV at a geomagnetic latitude of 52° (rigidity cut off $P_c = 2$GV). The dashed lines are from mid-latitude balloon observations by Harrison et al.[11] and Rosen and Hofmann[12] in (a) and (b), respectively, both having been normalised to $\phi = 270$ MV using the monthly mean $\phi$ at the time of the flight. The typical DLW profile was measured during a balloon flight by Philipona et al.[4]

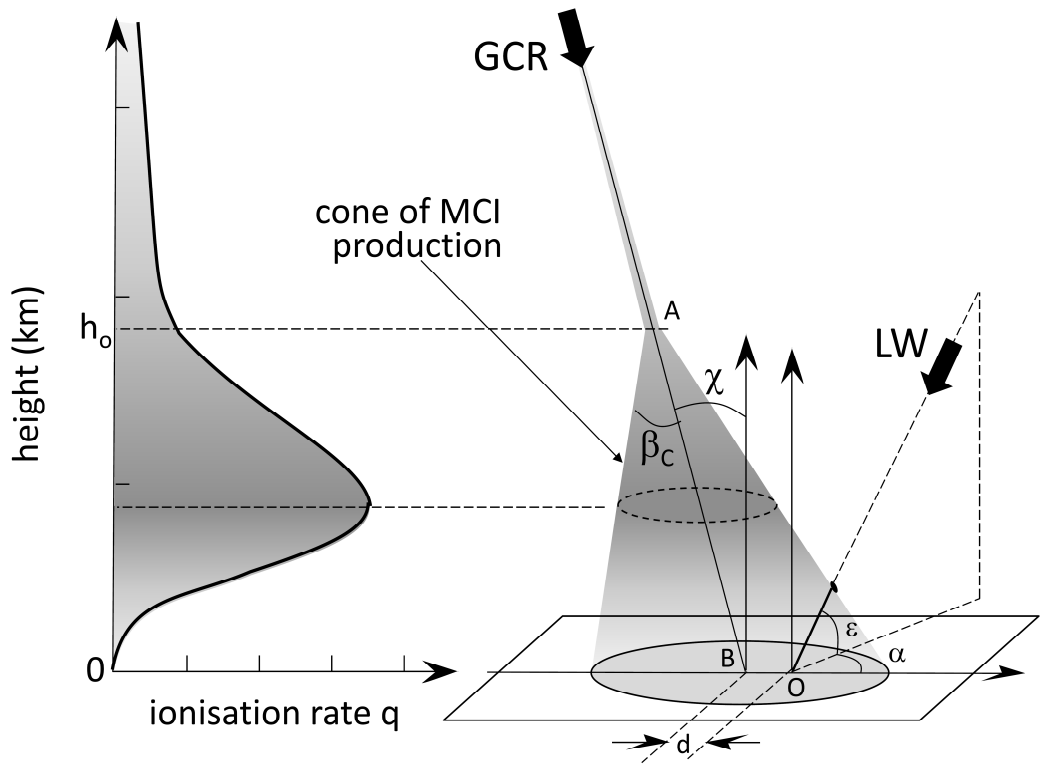

**Figure 2.** (left) Typical ion profile generated by an energetic primary GCR showing ion concentration, $N_i$, as a function of height, h. (right) Schematic of the resulting cone of ionisation (shaded in grey) centred on the point B on the ground, a distance d from the cosmic ray telescope and LW radiometer which are located at O. The size of the cone on the ground is set by the cone angle $\beta_C$ and the height of the top of the cone, $h_o$. The primary GCR precipitates at the zenith angle $\chi$. The elevation and azimuth angles of the radiometer field of view are $\varepsilon$ and $\alpha$, respectively.

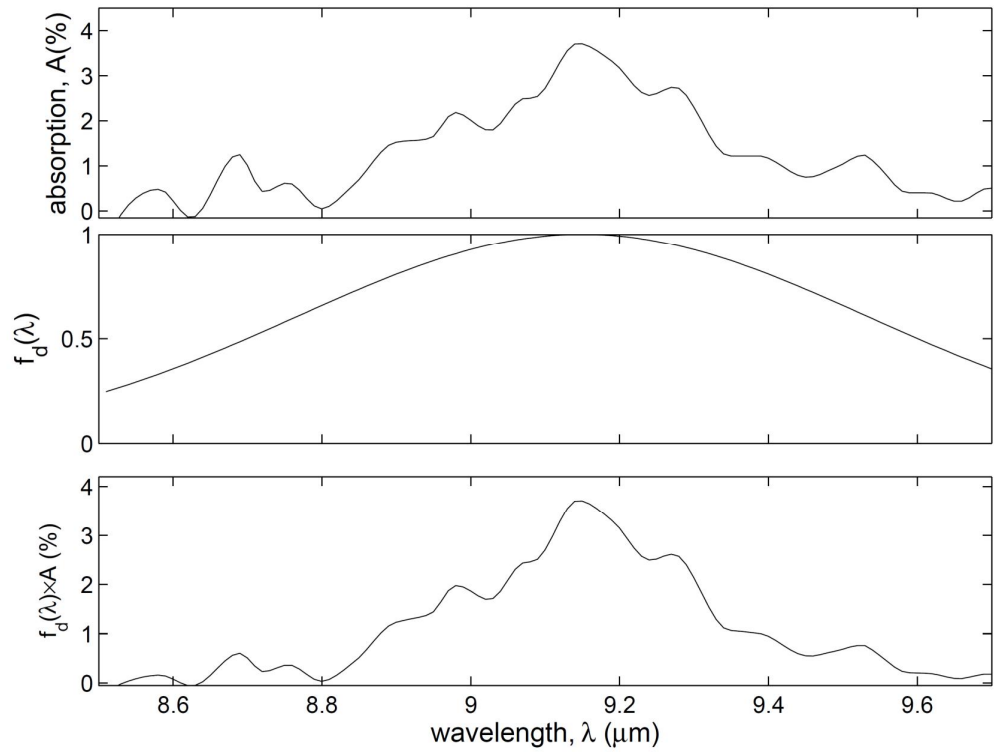

**Figure 3.** The spectrum of the ion absorption line and the narrowband radiometer response in the AL experiment. (Top panel) the absorption spectrum, $A(\lambda)$, observed in the laboratory.[6] (Middle panel) the narrowband filter response in the experiment of AL, $f_d(\lambda)$.[19] (Bottom panel) $A(\lambda) \times f_d(\lambda)$.

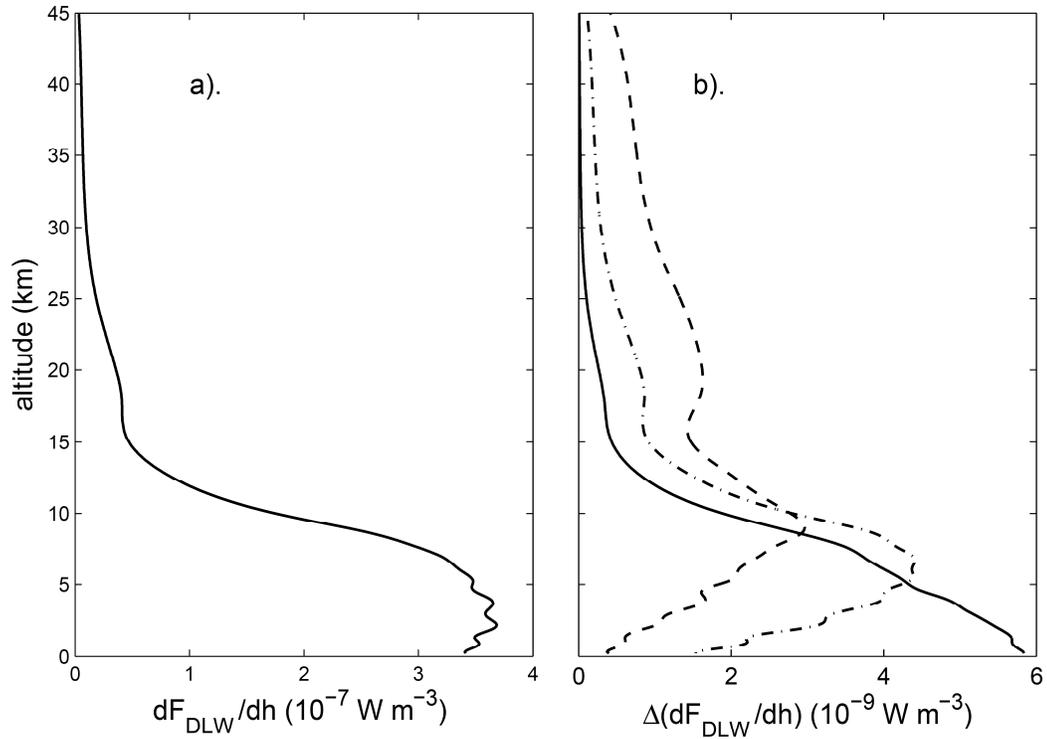

**Figure 4**. Profiles of the narrowband absorption per unit height. (a) Is the total for the steady state conditions shown in figure 1. (b) shows the changes in the profile compared to the steady state case introduced at the peak change by the three scenarios discussed in the text. The solid line is for scenario A, a 6% increase in $J_i(E)$ in the energy range 10<E<100GeV (for both protons and alpha particles); the dot-dash line is for scenario B, a 20% fall in the heliospheric modulation potential, $\phi$; and the dashed line for scenario C, a 20% fall in the geomagnetic cut-off rigidity, $P_C$.

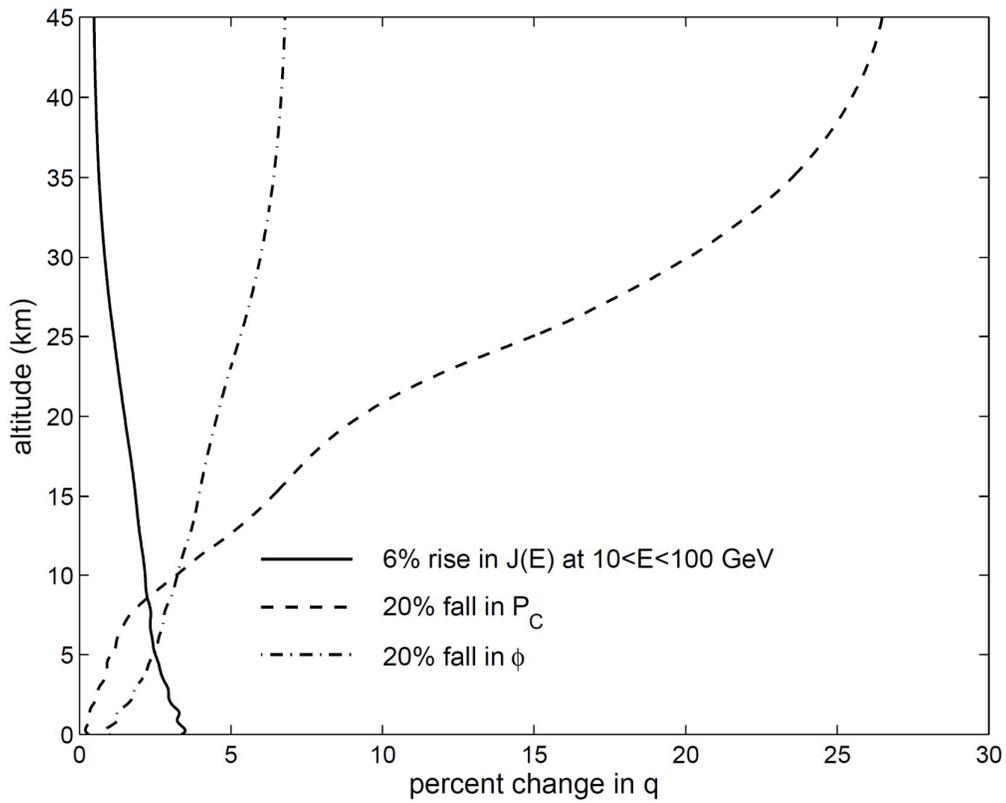

**Figure 5**. The percentage difference between the ionisation rate profiles for peak change and the pre-existing steady state for the three scenarios outline in the text. The line types are for the same scenarios as for figure 4(b): the solid line is for scenario A, a 6% increase in $J_i(E)$ in the energy range 10<E<100GeV (for both protons and alpha particles); the dot-dash line is for scenario B, a 20% fall in the heliospheric modulation potential, $\phi$; and the dashed line for scenario C, a 20% fall in the geomagnetic cut-off rigidity, $P_C$

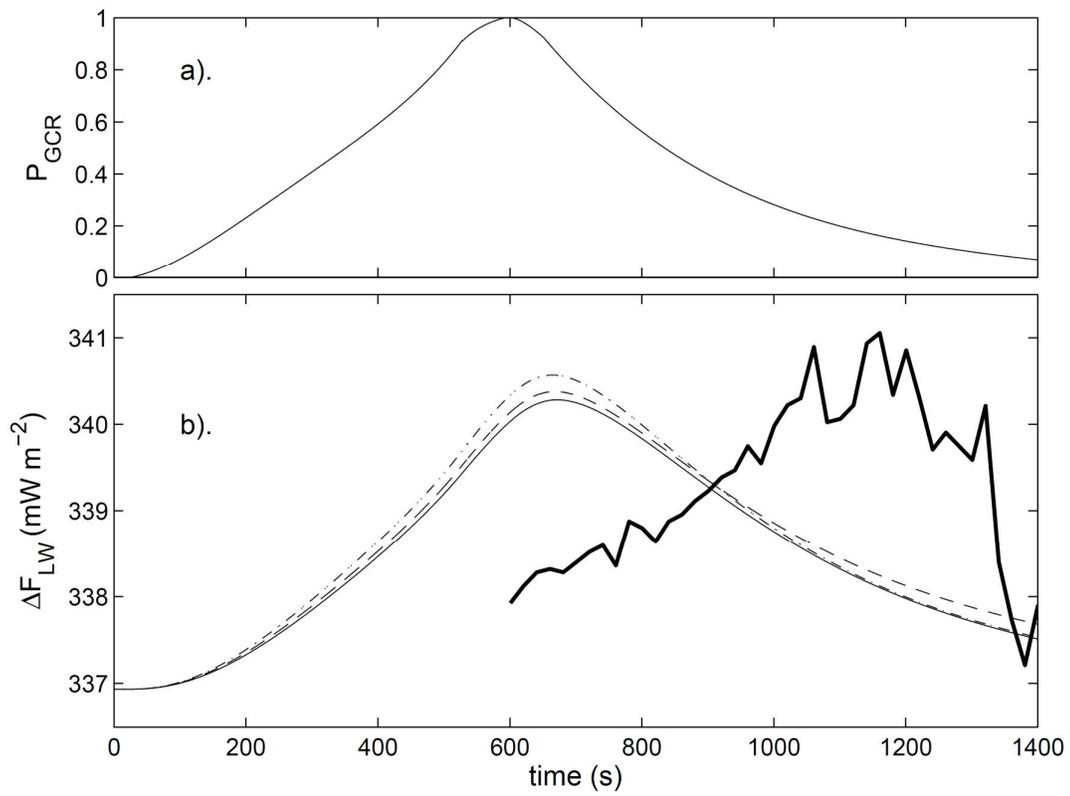

**Figure 6**. (a) The probability of coincident GCR detections as a function of time elapsed since each trigger event used in AL's composite analysis. (b) The modelled variations in DLW absorption ($\Delta F_{LW}$ is defined as positive for power absorbed) for the three scenarios (plotted using the same line types as used in figures 4 and 5). The thick solid line is the mean from AL's composite analysis (their figure 3).

**References**

[1] Erlykin A.D., T. Sloan and A.W. Wolfendale (2014), Cosmic rays and changes in atmospheric infra-red transmission, Astroparticle Physics, **57/58**, 26–29, doi: 10.1016/j.astropartphys.2014.03.007

[2] Aplin K.L. and M. Lockwood (2013), Cosmic ray modulation of infra-red radiation in the atmosphere, *Env Res Letts*, **8**, 015026 (6pp), doi:10.1088/1748-9326/8/1/015026

[3] Trenberth, K.E., J.T. Fasullo, and, J. Kiehl (2009), Earth's global energy budget, Bulletin of the American Meteorological Society, **90** (3), 311-323, doi: 10.1175/2008BAMS2634.1

[4] Philipona, R., A. Kräuchi, and E. Brocard (2012) Solar and thermal radiation profiles and radiative forcing measured through the atmosphere, Geophys. Res. Lett., **39**, L13806, doi: 10.1029/2012GL052087

[5] Carlon, H.R. (1982) Infrared absorption and ion content of moist atmospheric air, Infrared Physics, **22**, 43–49, doi:10.1016/0020-0891(82)90017-3

[6] Aplin K.L. and R.A. McPheat (2005) Absorption of infra-red radiation by atmospheric molecular cluster ions, J. Atmos. Sol-Terr. Phys., **67,** 775–83, doi: 10.1016/j.jastp.2005.01.007

[7] Lockwood, M. (2012) Solar Influence on Global and Regional Climate, Surveys in Geophysics, **33** (3), 503-534, doi: 10.1007/s10712-012-9181-3

[8] Jones, G.S., M. Lockwood, and P.A. Stott (2012) What influence will future solar activity changes over the 21st century have on projected global near surface temperature changes?J. Geophys. Res. (Atmos.), **117**, D05103, doi:10.1029/2011JD017013.

[9] Forster, P., V. Ramaswamy, P. Artaxo, T. Berntsen, R. Betts, D.W. Fahey, J. Haywood, J. Lean, D.C. Lowe, G. Myhre, J. Nganga, R. Prinn, G. Raga, M. Schulz and R. Van Dorland (2007) Changes in Atmospheric Constituents and in Radiative Forcing, *in Climate Change 2007: The Physical Science Basis. Contribution of Working Group I to the Fourth Assessment Report of the Intergovernmental Panel on Climate Change* [Solomon, S., D. Qin, M. Manning, Z. Chen, M. Marquis, K.B. Averyt, M.Tignor and H.L. Miller (eds.)]. Cambridge University Press, Cambridge, United Kingdom and New York, NY, USA. http://www.ipcc.ch/pdf/assessment-report/ar4/wg1/ar4-wg1-chapter2.pdf

[10] I G Usoskin and G Kovaltsov (2006), Cosmic ray induced ionization in the atmosphere: Full modeling and practical applications, J. Geophys. Res., **111**, #D21206, doi: 10.1029/2006JD007150

[11] Harrison R.G., Nicoll K.A. and Aplin K.L. (2014), Vertical profile measurements of lower troposphere ionisation, J. Atmos. Solar-Terr. Phys., **119**, 203-210 doi:10.1016/j.jastp.2014.08.006

[12] Rosen, J. M., and D. J. Hofmann (1981), Balloon-borne measurements of the small ion concentration, J. Geophys. Res., **86** (C8), 7399–7405, doi:10.1029/JC086iC08p07399

[13] Barnard, L., M. Lockwood, M.A. Hapgood, M.J. Owens, C.J. Davis, and F. Steinhilber (2011) Predicting Space Climate Change, Geophys. Res. Lett., **38**, L16103, doi:10.1029/2011GL048489

[14] Bazilevskaya, G.A., I.G. Usoskin, E.O. Flückiger, R.G. Harrison, L. Desorgher, R. Bütikofer, M.B. Krainev, V.S. Makhmutov, Y.I. Stozhkov, A.K. Svirzhevskaya, N.S. Svirzhevsky, and G.A. Kovaltsov (2008) Cosmic Ray Induced Ion Production in the Atmosphere, Space Sci. Rev., **137**, 149–173, doi: 10.1007/s11214-008-9339-y

[15] Usoskin, I.G., O.G. Gladysheva, and G.A. Kovaltsov (2004) Cosmic ray-induced ionization in the atmosphere: spatial and temporal changes, J. Atmos. Sol.-Terr. Phys., **66**, 1791–1796, doi: 10.1016/j.jastp.2004.07.037

[16] Walden, V. P., S. G. Warren, and F. J. Murcray (1998), Measurements of the downward longwave radiation spectrum over the Antarctic Plateau and comparisons with a line-by-line radiative transfer model for clear skies, J. Geophys. Res., 103(D4), 3825–3846, doi:10.1029/97JD02433.

[17] Forster, L., C. Emde, B. Mayer, and S. Unterstrasser (2012) Effects of three-Dimensional photon transport on the radiative forcing of realistic contrails, J. Atmos. Sci., **69**, 2243–2255, doi: 10.1175/JAS-D-11-0206.1

[18] Zeigler, J.F. (1998) Terrestrial cosmic ray intensities, IBM Journal of Research and Development, **42** (1), 117 – 140, doi: 10.1147/rd.421.0117

[19] Aplin, K.L. and R.A. McPheat (2008) An infrared filter radiometer for atmospheric cluster-ion detection, Rev. Sci. Inst., **79**, article # 106107, doi: 10.1063/1.3002428

[20] Clem, J. M., J. W. Bieber, P. Evenson, D. Hall, J. E. Humble, and M. Duldig (1997), Contribution of obliquely incident particles to neutron monitor counting rate, J. Geophys. Res., **102** (A12), 26919–26926, doi:10.1029/97JA02366

[21] Harrison R G and Carslaw K S (2003) Ion-aerosol-cloud processes in the lower atmosphere *Rev. Geophys*. **41,** 1012

[22] Bates, D.R. (1982) Recombination of small ions in the troposphere and lower stratosphere, Planet. Space Sci., **30**, (12), 1275–1282, doi: 10.1016/0032-0633(82)90101-5

[23] Rosen, J.M., and D.J. Hofmann (1981), Balloon-borne measurements of electrical conductivity, mobility, and the recombination coefficient, J. Geophys. Res., 86 (C8), 7406–7410, doi:10.1029/JC086iC08p07406.

[24] Thakur, N., and the BESS-Polar consortium (2013) Short-term Variations in Cosmic Ray Proton Fluxes from BESS-Polar I., in ICRC 2013, Proc. 33rd International Cosmic Ray Conference, Rio De Janeiro, 2-9 July, 2013, ed. A. Saa, http://www.cbpf.br/~icrc2013/papers/icrc2013-1088.pdf

[25] Rouillard, A,P. and M. Lockwood (2007) The latitudinal effect of co-rotating interaction regions on galactic cosmic rays, Sol. Phys., **245**, 191-206, doi: 10.1007/s11207-007-9019-1, 2007

[26] Thomas, S.R., M.J. Owens, M. Lockwood and C.J. Scott (2014) Galactic Cosmic Ray Modulation near the Heliospheric Current Sheet, Sol. Phys., **289** (7), 2653-2668, doi: 10.1007/s11207-014-0493

[27] Leske, R. A., R. A. Mewaldt, E. C. Stone, and T. T. vonRosenvinge (2001), Observations of geomagnetic cutoff variations during solar energetic particle events and implications for the radiation environment at the Space Station, J. Geophys. Res., **106** (A12), 30011–30022, doi:10.1029/2000JA000212.

[28] Harrison, R.G. (2000) Cloud Formation and the Possible Significance of Charge for Atmospheric Condensation and Ice Nuclei, Space Science Reviews, **94** (1-2), 381-396, doi: 10.1023/A:1026708415235

[29] Arnold F., D. Krankowsky and K. H. Marien, First mass spectrometric measurements of positive ions in the stratosphere, Nature **267**, 30-32 (1977) doi:10.1038/267030a0doi:10.1038/267030a0

[30] Goss L.M.,S.W. Sharpe, T.A. Blake, V. Vaida, and J.W. Brault (1999) Direct Absorption Spectroscopy of Water Clusters, J. Phys. Chem. A, **103** (43), pp 8620–8624, doi: 10.1021/jp9920702

[31] Dunn, M. E., Pokon, E. K., & Shields, G. C. (2004). Thermodynamics of forming water clusters at various temperatures and pressures by Gaussian-2, Gaussian-3, complete basis set-QB3, and complete basis set-APNO model chemistries; implications for atmospheric chemistry. J. Am. Chem. Soc., 126(8), 2647-2653.

[32] Ndongmouo, U. F. T., Lee, M. S., Rousseau, R., Baletto, F., & Scandolo, S. (2007). Finite-temperature effects on the stability and infrared spectra of $HCl(H_2O)_6$ clusters. J. Phys. Chem. A, **111**(49), 12810-12815.

[33] Pierre Auger Collaboration. (2011). The Pierre Auger Observatory scaler mode for the study of solar activity modulation of galactic cosmic rays. Journal of Instrumentation, 6(01), P01003

[34] Acounis, S., Charrier, D., Garçon, T., Rivière, C., & Stassi, P. (2012). Results of a self-triggered prototype system for radio-detection of extensive air showers at the Pierre Auger Observatory. Journal of Instrumentation, 7(12), P11023.